\documentclass[aps,prl,twocolumn,showpacs,preprintnumbers,amsmath,amssymb,superscriptaddress]{revtex4-1}
\usepackage{graphicx}% Include figure files
\usepackage{dcolumn}% Align table columns on decimal point
\usepackage{bm}% bold math
\usepackage{hyperref}
\newcommand{\nc}{\newcommand}
\nc{\be}{\begin{equation}}
\nc{\ee}{\end{equation}}
\nc{\bea}{\begin{eqnarray}}
\nc{\eea}{\end{eqnarray}}
\nc{\bean}{\begin{eqnarray*}}
\nc{\eean}{\end{eqnarray*}}
\nc{\mb}{\mbox}
\nc{\rnc}{\renewcommand}
\nc{\vk}{\mb{\bmk}}
\nc{\vp}{\mb{\bmp}}
\nc{\vn}{\mb{\bmn}}
\nc{\vq}{\mb{\bmq}}
\nc{\rr}{\mb{\bmr}}
\nc{\vz}{\hat {\mb{\bmz}}}
\nc{\vj}{\mb{\boldmath$j$}}
\nc{\vg}{\mb{\boldmath$g$}}
\nc{\x}{\mb{\boldmath$x$}}
\nc{\A}{\mb{\boldmath$A$}}
\nc{\va}{\mb{\boldmath$a$}}
\nc{\vs}{\mb{\boldmath$\sigma$}}
\nc{\vpi}{\mb{\boldmath$\pi$}}
\nc{\nab}{\nabla}
\nc{\X}{\sf x}
\usepackage{txfonts}
\usepackage{pxfonts}
\usepackage{graphicx}
\usepackage{lipsum}

\begin{document}

\title{Quantum Parity Hall effect in ABA Graphene}

\author{Petr Stepanov}
\thanks{These authors contributed equally.}
\affiliation{Department of Physics and Astronomy, Ohio State University, Columbus, OH 43220}

\author{Yafis Barlas} 
\thanks{These authors contributed equally.}
\affiliation{Department of Physics, Yeshiva University, New York, NY, USA}

\author{Shi Che}
\affiliation{Department of Physics and Astronomy, Ohio State University, Columbus, OH 43220}
\author{Kevin Myhro}
\affiliation{Department of Physics and Astronomy, University of California, Riverside, CA 92521}
\author{Greyson Voigt}
\affiliation{Department of Physics and Astronomy, University of California, Riverside, CA 92521}
\author{Ziqi Pi}
\affiliation{Department of Physics and Astronomy, University of California, Riverside, CA 92521}
\author{Kenji Watanabe} 
\affiliation{National Institute for Materials Science, 1-1 Namiki Tsukuba Ibaraki 305-0044 Japan.}
\author{Takashi Taniguchi} 
\affiliation{National Institute for Materials Science, 1-1 Namiki Tsukuba Ibaraki 305-0044 Japan.}
\author{Dmitry Smirnov}
\affiliation{National High Magnetic Field Laboratory, Tallahassee, FL 32310.}
\author{Fan Zhang} 
\affiliation{Department of Physics, University of Texas at Dallas, Richardson, TX 75080.}
\author{R. Lake}
\affiliation{Department of Electrical Engineering, University of California, Riverside, CA 92521}
\author{Allan MacDonald}
\affiliation{Department of Physics, University of Texas at Austin, Austin, TX 78712-1192.}
\author{Chun Ning Lau} 
\affiliation{Department of Physics and Astronomy, University of California, Riverside, CA 92521}
\affiliation{Department of Physics and Astronomy, Ohio State University, Columbus, OH 43220}

% Please give the surname of the lead author for the running footer

% Please include corresponding author, author contribution and author declaration information
%\authorcontributions{P.S. and C.N.L. conceived the project. K.W. and T.T. provided materials. P.S., S.C., K.M., G.V. and Z.P. fabricated samples. P.S., S.C. and D.S. performed measurements. P.S., C.N.L. and Y.B. analyzed and interpreted data. Y.B., F.Z., R.L. and A.H.M. performed theoretical analysis. P.S., Y.B., C.N.L. and A.H.M. wrote the manuscript. All authors discussed and commented on the manuscript.}

%\authordeclaration{The authors declare no conflict of interest.}
%\equalauthors{\textsuperscript{1}P.S (Petr Stepanov) and Y.B. (Yafis Barlas) contributed equally to this work.}
%\correspondingauthor{\textsuperscript{2}To whom correspondence should be addressed: macdpc@physics.utexas.edu and lau.232@osu.edu }

% Keywords are not mandatory, but authors are strongly encouraged to provide them. If provided, please include two to five keywords, separated by the pipe symbol, e.g:
%\keywords{Topological materials $|$ Dynamical systems $|$ Symmetry $|$} 

%\begin{abstract}
%Please provide an abstract of no more than 250 words in a single paragraph. Abstracts should explain to the general reader the major contributions of the article. References in the abstract must be cited in full within the abstract itself and cited in the text.
%\end{abstract}
\begin{abstract}
The celebrated phenomenon of quantum Hall effect has recently been generalized from transport of conserved charges to that of other approximately conserved state variables, including spin and valley~\cite{ref1,ref2,ref3,ref4,ref5,ref6,ref7,ref8}, which are characterized by spin- or valley-polarized boundary states with different chiralities. Here, we report a new class of quantum Hall effect in ABA-stacked graphene trilayers (TLG), the quantum parity Hall (QPH) effect, in which boundary channels are distinguished by even or odd parity under the system’s mirror reflection symmetry. At the charge neutrality point and a small perpendicular magnetic field $B_{\perp}$, the longitudinal conductance $\sigma_{xx}$ is first quantized to $4e^2/h$, establishing the presence of four edge channels. As $B_{\perp}$ increases, $\sigma_{xx}$ first decreases to $2e^2/h$, indicating spin-polarized counter-propagating edge states, and then to approximately $0$. These behaviors arise from level crossings between even and odd parity bulk Landau levels, driven by exchange interactions with the underlying Fermi sea, which favor an ordinary insulator ground state in the strong $B_{\perp}$ limit, and a spin-polarized state at intermediate fields. The transitions between spin-polarized and unpolarized states can be tuned by varying Zeeman energy. Our findings demonstrate a topological phase that is protected by a gate-controllable symmetry and sensitive to Coulomb interactions.  
\end{abstract}

%\dates{This manuscript was compiled on \today}
%\doi{\url{www.pnas.org/cgi/doi/10.1073/pnas.XXXXXXXXXX}}

% Optional adjustment to line up main text (after abstract) of first page with line numbers, when using both lineno and twocolumn options.
% You should only change this length when you've finalised the article contents.
%\verticaladjustment{-2pt}

\maketitle
%\thispagestyle{firststyle}
%\ifthenelse{\boolean{shortarticle}}{\ifthenelse{\boolean{singlecolumn}}{\abscontentformatted}{\abscontent}}

% If your first paragraph (i.e. with the \dropcap) contains a list environment (quote, quotation, theorem, definition, enumerate, itemize...), the line after the list may have some extra indentation. If this is the case, add \parshape=0 to the end of the list environment.

In the conventional Hall effect, a charge current combines with a perpendicular magnetic field $B_{\perp}$ to yield a steady state with a transverse chemical potential gradient.  A quantum version of the Hall effect (QHE), in which the chemical potential gradient is replaced by a chemical potential difference between opposite sample edges, can occur when the two-dimensional bulk is insulating.  Recently, the Hall effect and the QHE have been generalized from transport of conserved charge to transport of other approximately conserved state variables, including spin~\cite{ref1,ref2} and valley~\cite{ref3, ref4}; their quantum versions~\cite{ref5,ref6,ref7,ref8} are then characterized by spin or valley polarized boundary states with different chiralities. For material systems that host these topologically non-trivial phenomena, discrete symmetries play an important role. For example, time reversal symmetry in the quantum spin Hall effect is essential to protect one-dimensional counter-propagating edge modes from backscattering. 

\begin{widetext}

\begin{figure}
%\begin{center}
\includegraphics[width=0.9\linewidth]{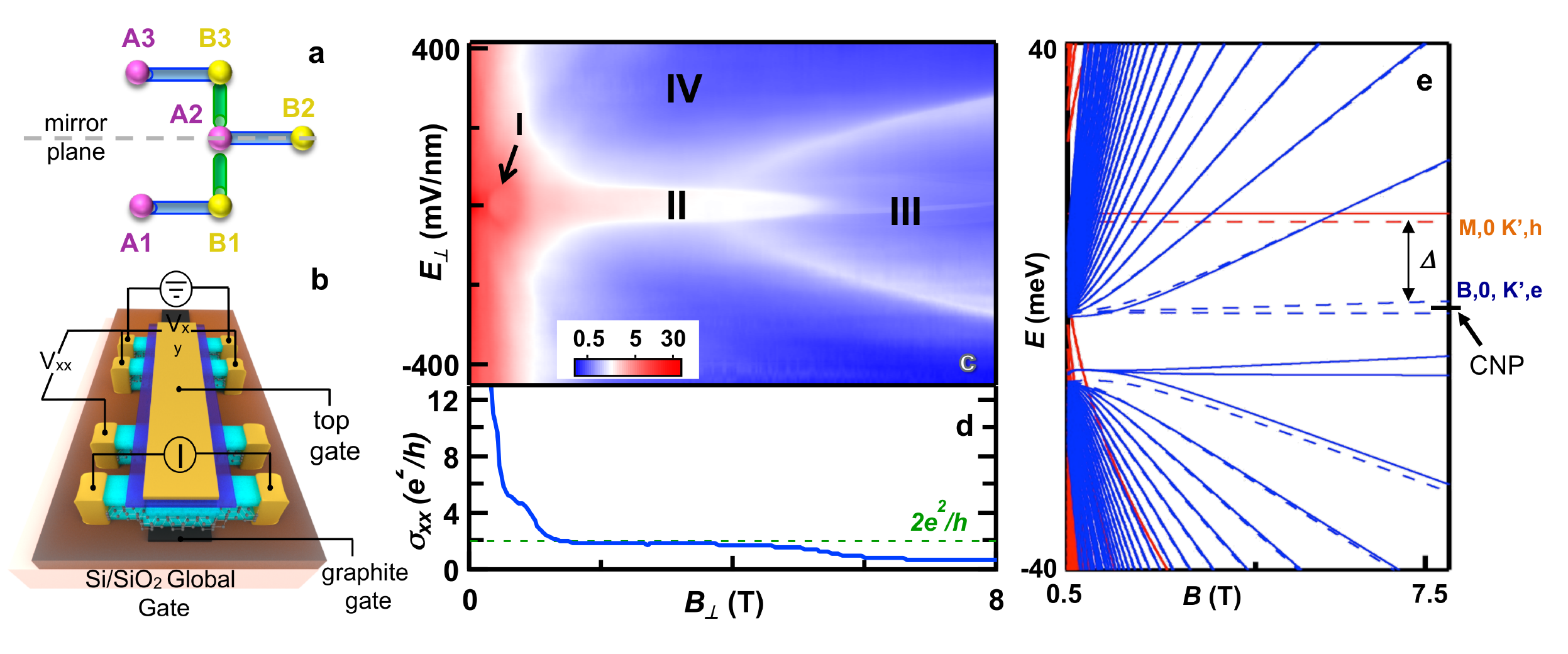}
\caption{(a). Atomic configurations of ABA TLG. (b). Schematics of hBN-encapsulated TLG device. (c). Phase diagram $\sigma_{xx}(E_{\perp}, B_{\perp}) $ at the CNP and $T=260$ mK. The different electronic phases are labeled I through IV. (d). Line trace $\sigma_{xx}(B_{\perp})$ at $E_{\perp}=0$. (e). Full parameter $k \cdot p$ model calculations of the LL energy spectrum as a function of the B field. Dashed (solid) lines indicate $K$ ($K$’) valley; red and blue lines indicate LL arising from MLG- and BLG-like branches that are odd and even under mirror reflection operation, respectively.}
\label{Figone}
%\end{center}
\end{figure}

\end{widetext}

Multi-band Dirac systems, such as ABA-stacked trilayer graphene (TLG), afford richer and more exotic symmetry-protected topological (SPT) phases. For instance, in addition to approximate spin and valley symmetries, TLG has an additional discrete mirror symmetry (Fig. 
~\ref{Figone} A), that allows bands to be classified by their parity, the eigenvalue of the operator for reflection in the plane of the middle layer.  TLG has two low-energy odd-parity $\pi$-bands described by a massless Dirac Hamiltonian with a monolayer graphene (MLG) -like spectrum, and four low-energy even parity bands that exhibit a massive Dirac bilayer graphene (BLG) -like spectrum~\cite{ref9,ref10,ref11,ref12,ref13,ref14,ref15,ref16,ref17,ref18,ref19,ref20,ref21,ref22,ref23,ref24}. As long as mirror symmetry is preserved, Landau levels (LLs) belonging to different mirror symmetry representations do not couple. 

The complex interplay between the spin, valley, parity, and electronic interactions in ABA graphene suggests the possibility of symmetry-protected topological phases at the carrier neutrality point (CNP) that have not been identified in previous studies~\cite{ref25,ref26,ref27,ref28,ref29,ref30}.  In this paper, we demonstrate SPT phases in ABA-trilayer graphene in which mirror symmetry preserves counter-propagating edge modes. We observe a sequence of transitions between different symmetry protected topological phases that are driven by magnetic-field dependent interactions between electrons close to the Fermi level and the Dirac sea. When mirror symmetry is broken by an out-of-plane displacement field, the two-probe longitudinal conductance decreases dramatically to form a layer polarized insulator.

Our experiments were performed on dual-gated TLG devices encapsulated between two hBN sheets (Fig.~\ref{Figone} B), etched into Hall-bar geometries and with edge contacts~\cite{ref31,ref32}. The device quality was enhanced by the presence of a local graphite gate underneath the TLG channel, which provides for additional screening of charged impurities and also enables independent control of the Fermi levels of the channel and the leads~\cite{ref33}. The device is very conductive in the absence of external fields, with longitudinal conductivity $\sigma_{xx} \sim $ 1mS at the CNP, with estimated quantum mobility of $\sim 80,000 \rm{cm}^2/(\rm{V}.\rm{s})$ ~\cite{ref24}. By varying top ($V_{TG}$) and bottom gates ($V_{BG}$), we independently control the charge carrier density n and the external displacement field ($E_{\perp}$). In TLG, $E_{\perp}$ breaks the mirror symmetry of the ABA film, which allows us to explore the symmetries associated with the topological phases. 

	To explore topological phases at the CNP, we measure the device’s $\sigma_{xx}$ as a function of $B_{\perp}$ and $E_{\perp}$ while maintaining $N=0$. The resulting phase diagram is shown in Fig.~\ref{Figone} C. Strikingly, at least four different phases can be identified, labeled as I through IV on the plot. The first three phases are observed near zero $E_{\perp}$: phase I with quantized conductance $\sigma_{xx}\sim 4e^2/h$ occurs for very small fields, $\sim 0.5$ T; phase II has half the conductance of phase I ($\sigma_{xx} \sim 2e^2/h$) and emerges at intermediate strengths of $B_{\perp}$ ($1 < B_{\perp} < 4$T) (Fig.~\ref{Figone} D). As $B_{\perp}$ increases further, the device transitions to phase III, a resistive state with measured $\sigma_{xx} \sim 0.5 e^2/h$. In contrast to the first 2 phases, phase III persists over a wide range in $|E_{\perp}|$. All 3 phases are destroyed by the application of sufficiently large $|E_{\perp}|$, as the device transitions into an insulating phase IV with $\sigma_{xx} <0.1 e^2/h$.

	This striking phase diagram identifies different phases by their longitudinal conductivity values and points to symmetry-protected topological phases at the CNP. To understand the nature of these states, we first calculate the band structure of TLG using a continuum $k \cdot p$ model. The various remote hopping parameters were extracted by fitting calculated spectra to LL crossings in the experimental Landau fan diagram, and are consistent with previous work~\cite{ref25,ref28,ref31,ref35}. The calculated LL spectrum is displayed in Fig.~\ref{Figone} E, in which even parity BLG-like (B) LLs are represented by blue lines, odd parity MLG-like (M) branches by red lines, and LLs associated with K and K’ valley LLs are plotted as solid and dashed lines respectively. A particularly important feature of ABA graphene is the energetic displacement between the even- and odd-parity Dirac points, which arises from remote interlayer hopping and energy differences between stacked and unstacked atoms.  In B, this displacement places the four-fold degenerate lowest LLs in the MLG-like branch well above the 8-fold degenerate lowest LLs in the BLG-like branch. As shown in Fig.~\ref{Figone} E, the non-interacting electron ground state for $B_{\perp} \gtrapprox 0.5 T$ is a QPH state with LL filling factor $\nu=-2$ for odd parity states and $\nu=+2$ for even parity states. As explained below, the interplay between parity, Zeeman energy, and interactions in these two LLs explains most of the phase diagram.

\begin{figure}
\begin{center}
\includegraphics[width=0.9\linewidth]{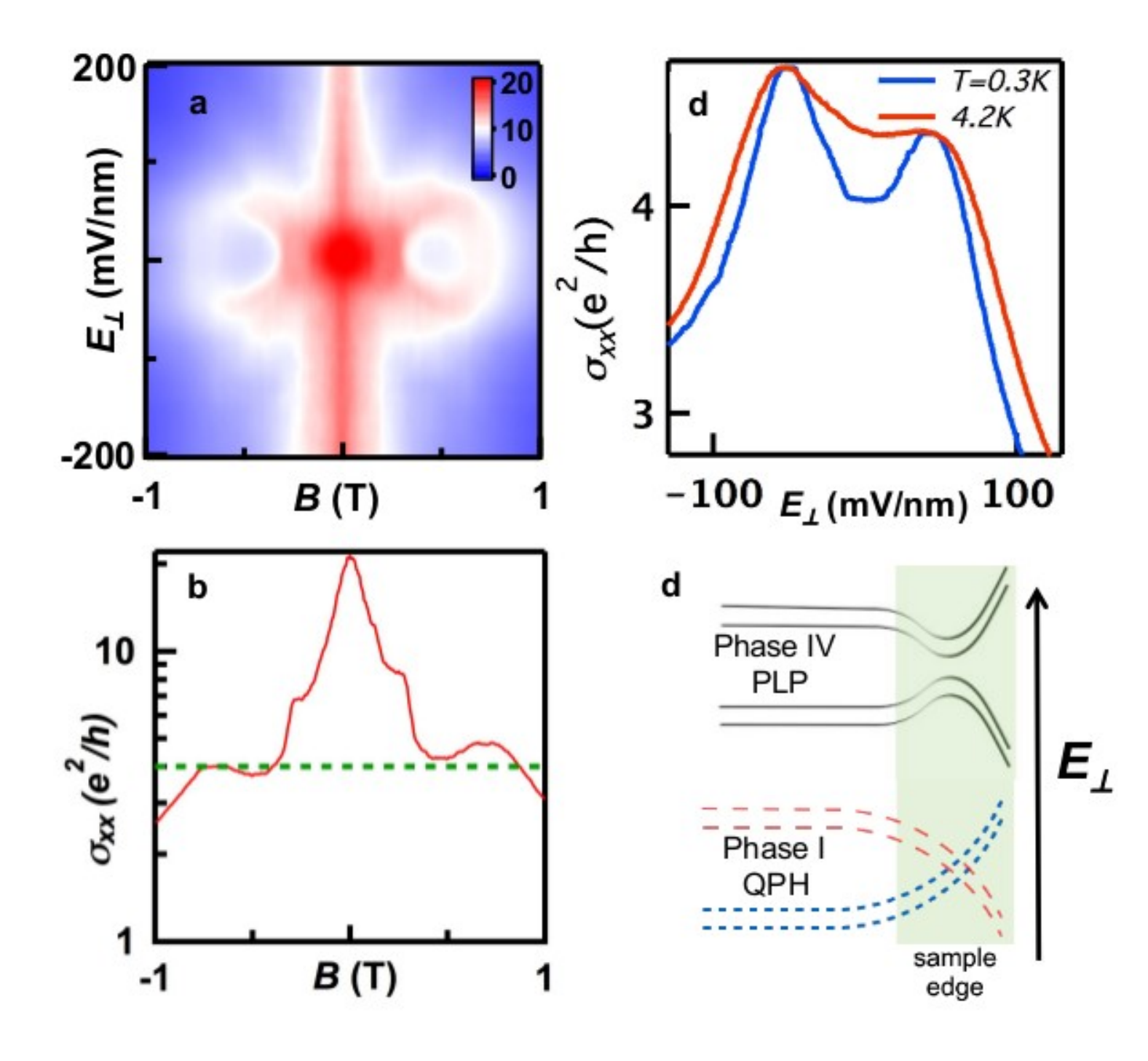}
\caption{Phase I of in TLG. (a). High resolution $\sigma_{xx}(E_{\perp}, B_{\perp}) $ at low magnetic field $|B_{\perp}| < 1$. (b) $\sigma_{xx}(B_{\perp}) $ at $ E_{\perp} = 0$, showing two plateaus $\sim 4e^2/h$ at $B_{\perp} \sim ±0.5T$. (c) $\sigma_{xx}(E_{\perp}) $ at $B_{\perp}=-0.5T$ . Blue and red lines are taken at $260$ mK and $4.2$ K, respectively. (d). Schematics of counter-propagating edge states consisting of $0- $ LL in MLG-like band and $0+$ in the BLG-like band. The lower and upper schematics illustrate the edge states at zero $E_{\perp}$ and large $E_{\perp}$, respectively.}
\label{Figtwo}
\end{center}
\end{figure}

We first examine Phase I, which appears as a small “pocket” for $0.4<|B_{\perp}| < 0.8$T around $E_{\perp}=0$ (Fig.~\ref{Figtwo} A); within the pocket,$\sigma_{xx}$ displays a persistent plateau $\sim 4e^2/h$ (Figures 2B-C), while the Hall conductivity $\sigma_{xy} \sim 0$, demonstrating ballistic conduction along $4$ edge channels. The $\sigma_{xx}$ plateau is destroyed by application of $|E_{\perp}| > 50$ mV/nm, demonstrating that it is protected by mirror symmetry. These two properties identify Phase I as the spin-unpolarized non-interacting QPH effect ground state expected on the basis of single-particle physics. The quantized $\sigma_{xx}$ is associated with two  even-parity and two odd-parity chiral edge states that propagate from source to drain along opposite edges (Fig.~\ref{Figtwo} D). These approximately spin-degenerate counter-propagating edge states are protected against backscattering by an underlying crystalline symmetry, since they correspond to different representations of the mirror reflection symmetry of the TLG crystal lattice that is preserved at $E_{\perp}=0$. (The slight deviation of $\sigma_{xx}$ from the quantized value may be accounted for by the presence of disorder that breaks mirror symmetry). In the presence of a finite displacement field, backscattering between counter propagating states is allowed, the conductance is quickly reduced to a small value, and the ground state is a partially layer polarized ordinary insulator.

\begin{figure}
\begin{center}
\includegraphics[width=0.9\linewidth]{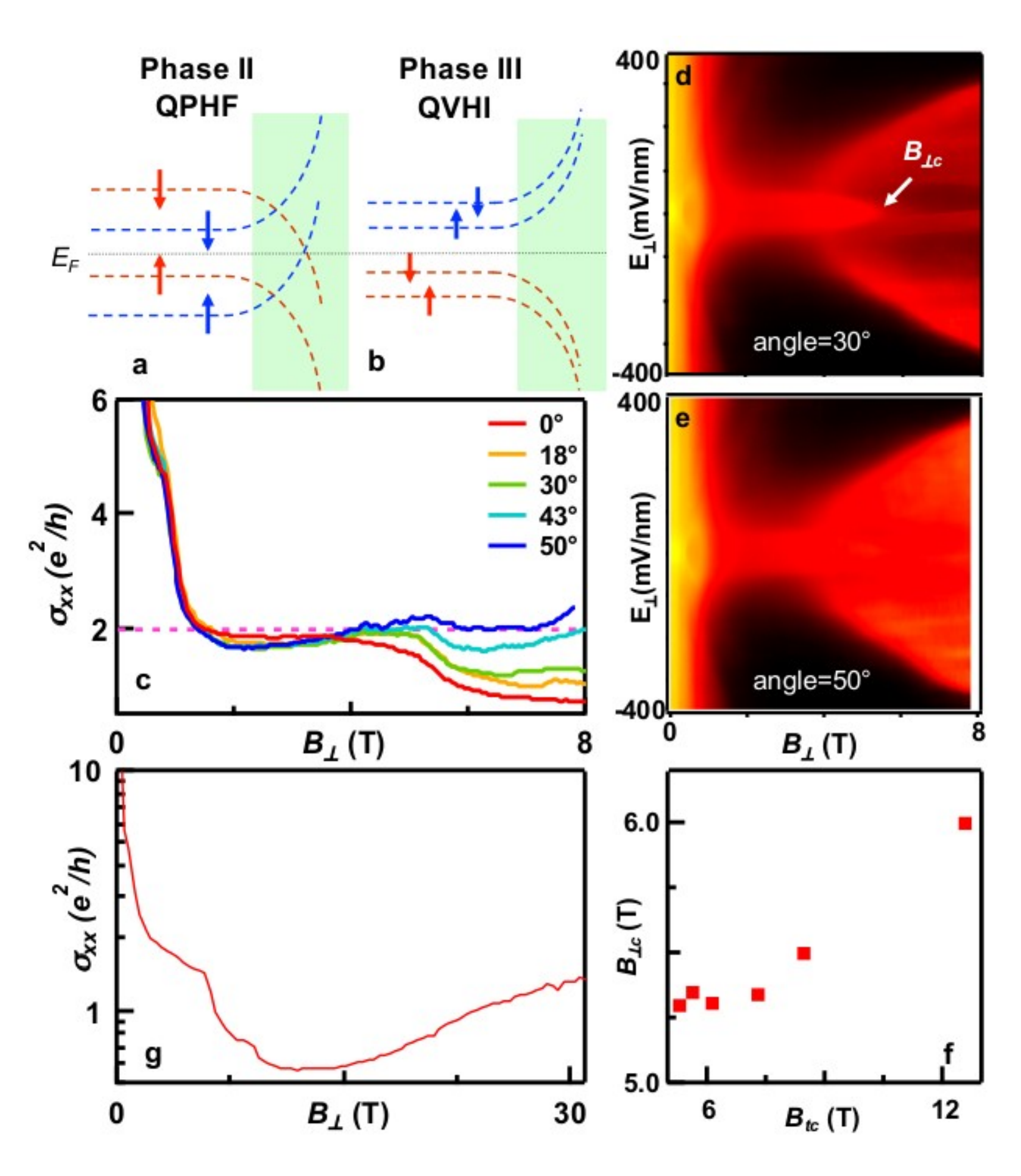}
\caption{Phases II and III of TLG. (a-b). Schematics of edge state configurations for Phase II and III, respectively. (c) $\sigma_{xx}(E_{\perp}, B_{\perp}) $  at different tilt angles. (d-e). $\sigma_{xx}(E_{\perp}, B_{\perp}) $ at tilt angles $\theta = 30^{o}$ and $50^{o}$ respectively. Color scale: black: low conductance; red: high conductance. (f). Critical field $B_{\perp,c}$ for transition between Phases II and III vs. total magnetic field $B_{t}$. (g). High field dependence $\sigma_{xx}( B_{\perp}) $ at $\theta=0$ for $B_{\perp}$ up to $31$T.}
\label{Figthree}
\end{center}
\end{figure}

 As $B_{\perp}$ increases from $1$T to $5$T, $\sigma_{xx}$ drops to $ \sim 2e^2/h$, signaling a reduction in the number of chiral channels from 4 in phase I to 2 in phase II. (Fig.~\ref{Figone} D). This reduction in $\sigma_{xx}$ is not expected in a non-interacting electron theory, and can explained only by taking Coulomb interactions into consideration.  Two aspects play an essential role: i) intra-LL interactions along with Zeeman energy can stabilize strongly spin-polarized monolayer-like and bilayer-like $N=0$ Landau levels, and ii) exchange interactions between the $N=0$ Landau levels and Dirac sea that induce a magnetic-field dependent change in the energy separation between LLs with different parity, and in the BLG-like case, different orbital character. The latter effect, which also plays a role in Phase I by selecting $N=0$ BLG-like LLs for occupation over N=1 LLs, can be captured by adding self-energy corrections~\cite{ref35,ref36,ref37} to the electron and hole-LL energies that are larger for N=1 in the BLG-like case. These shifts, which are related to the established Dirac velocity enhancement in graphene monolayers in the absence of a field, have the same sign as the carrier, and therefore lower the MLG-like LL energies while raising the BLG-like LL energies (for detailed calculations, see~\cite{ref24}). For intermediate $B_{\perp}$, the (M, h, $\uparrow$) LL is occupied in preference to the orbital $N=0$ (B, e, $\downarrow$) LL (Fig.~\ref{Figthree} A). Phase II has $\nu=-1$ for odd parity because of an empty spin-$\downarrow$ LL and $\nu=+1$ for even parity because of an occupied spin-$\uparrow$ LL, and therefore has counter-propagating edge states with opposite spin and opposite parity, explaining its accurate conductance quantization. We classify phase II as a quantum parity Hall ferromagnet (QPHF). 
	
	The quantized conductances of both phase I and II are suppressed by application of a large $|E_{\perp}|$.  Because $|E_{\perp}|$ breaks mirror symmetry, edge conduction is no longer protected by crystal symmetry when $|E_{\perp}| \neq 0$.  Because the $|E_{\perp}| \neq 0$ suppression occurs to phase II, we conclude that spin-rotational invariance, which would otherwise protect ballistic conduction at finite 
$|E_{\perp}|$, must be broken by spin-orbit coupling in our samples.  Even with spin-orbit coupling, mirror symmetry protects the 
$|E_{\perp}|=0$ spin-polarized counter-propagating states. For particles with spin, $\hat{M}^2=-1$, where $\hat{M}$ is the mirror symmetry operator~\cite{ref24}, with eigenvalues $ \pm \imath$ for the even parity and $\mp \imath $ for the odd parity up-spin(down-spin) eigenstates. Therefore, in the QPHF phase the projected mirror symmetry operator $\bar{M}$ satisfies $\bar{M}^2=-1$, providing an effective Kramer degeneracy for the QPHF phase. When $|E_{\perp}| \neq 0$, the spin polarized edge states intermix due to spin-orbit coupling, which was originally forbidden due to the mirror symmetry. Since classification on the basis of mirror symmetry is not relevant at $|E_{\perp}| \neq 0$, we identify phase IV as a partially layer polarized (PLP) ordinary insulator state with no chiral edge channels and no QPH effect, as shown in the upper schematic in Fig.~\ref{Figtwo} D.

 	For even larger $B_{\perp}$, electronic interactions become stronger, further depressing the (M, h) while elevating the (B, e) LL energies. When both spin polarizations of the former are occupied and both spin polarizations of the latter are empty, there are no longer counter-propagating edge states, as all the even-parity electron LLs are above the odd-parity hole LLs (Fig.~\ref{Figthree} B). This leads to an insulating phase, phase III, with valley polarization within each Dirac band, but no parity polarization. Phase III is a spin unpolarized quantum valley ferromagnetic insulator (QVFI) which has $\sigma_{xx} \sim 0.5 e^2/h$; we attribute the finite $\sigma_{xx}$ to weak intervalley scattering and disorder.
          
Stability conditions for the $|E_{\perp}| = 0$ phases we observe can be expressed (24) in terms of dressed LL energies: 
\begin{eqnarray}
\Sigma \leq \Delta - E_{Z}		&	&  \qquad {\rm Phase I (QPH)} \\ 					
\Delta -E_{Z} \leq \Sigma  \geq \Delta+E_{Z}	& & \qquad 		{\rm Phase II (QPHF)} \\            
\Sigma \geq \Delta +E_Z		& &  \qquad {\rm Phase III (QVFI)}		
\end{eqnarray}
where $E_Z$ is the Zeeman energy, $\Sigma $ is the relevant self-energy sum related to interactions with the Dirac sea and $\Delta$ is the separation between the spin-degenerate lowest LLs between the even- and odd-parity branches in the zero-field limit. From fitting LL crossing points~\cite{ref27,ref30,ref34}, $\Delta $ is estimated to be 14.5 meV. Transitions between the phases occur at equality. Despite their apparent simplicity, these equations capture all the salient features of our experimental observations.

\begin{figure}
\begin{center}
\includegraphics[width=0.9\linewidth]{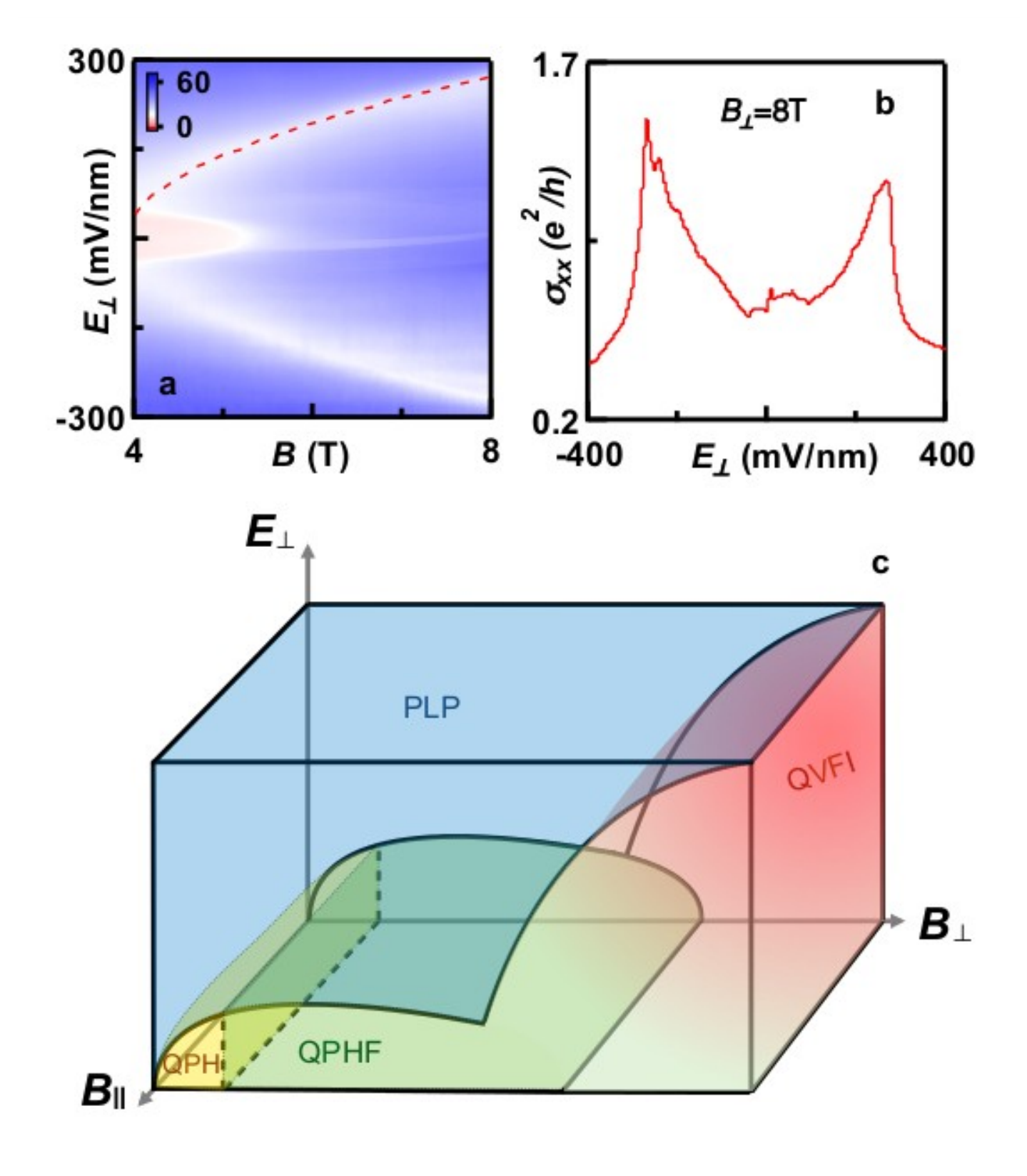}
\caption{Phase IV and overall phase diagram. (a) $\sigma_{xx}(E_{\perp}, B_{\perp}) $ at the tilt angle $\theta =0^{o}$ for $4 \leq B_{\perp} \geq 8$T. The dotted line is plotted  using $E_{\perp} ({\rm mV/nm})=135\sqrt{(B_{\perp}-4)}$. (b). Line trace $\sigma_{xx}(E_{\perp})$ at 
$B_{\perp} = 8 $T. (c). Schematics of the phase diagram of charge neutral TLG as a function of $E_{\perp}, B_{\perp} $ and $B_{||}$.}
\label{Figfour}
\end{center}
\end{figure}

Eq. (3) implies that the transition between phase II and phase III is tunable by the Zeeman energy. Adding an in-plane magnetic field $B_{||}$ increases the right-hand side of Eq. (3) while leaving $\Sigma$ which increases with $B_{||}$, unchanged.  It follows that phase III should be driven to phase II by a sufficiently large $B_{||}$. To test this prediction, we have measured $\sigma_{xx}$( $E_{\perp}$, $B_{\perp}$) at different magnetic field tilt angles $\theta$. The line traces $\sigma_{xx}( B_{\perp} )$ are shown in Fig.~\ref{Figthree} C, and two representative data sets at $\theta=30^{o}$ and $50^{o}$ are shown in Fig.~\ref{Figthree} D-E. As expected, as $B_{||}$ increases, the device crosses over to a conductive state that has $\sigma_{xx} \sim 2e^2/h$. Moreover, the critical $B_{\perp,c}$ value at which the transition to the QVFI state occurs increases with increasing $B_{||}$, in agreement with Eq. (3) (Fig.~\ref{Figthree} F).  We emphasize that, though superficially resembling the in-plane field phase transitions observed in MLG~\cite{ref38} and BLG~\cite{ref39}, this transition has a fundamentally different nature – instead of crossing over from a canted anti-ferromagnetic insulator to a ferromagnet in thinner graphene systems, TLG transitions from a spin unpolarized QVFI to a QPHF. Another important prediction of Eq. (3) is that since $\Sigma$ varies approximately like $\sqrt{B_{\perp}}$~\cite{ref24}, while $E_Z$ scales linearly with $B_{\perp}$, phase II should reemerge at fields stronger than we have discussed so far.  This is exactly what we observe as non-monotonic dependence of $\sigma_{xx}$ at very large $B_{\perp}$. As plotted in Fig.~\ref{Figthree} G, $\sigma_{xx}$ reaches a minimum at $12.5$T, then increase almost linearly as $B_{\perp}$ increases to 31T, as the device re-enters phase II.

	Finally, we find that at large $B_{\perp}$ the behavior of the conductivity as a function of displacement field is not monotonic.  As $E_{\perp}$ increases, $\sigma_{xx}$ first rises to $\sim 1.5 e^2/h$, then drops sharply to $\sim 0.1 e^2/h$ (Fig.~\ref{Figfour} B). The position in ($E_{\perp},B)$ space of the local conductivity peak that separates these two regimes can be described phenomenologically by the equation $E_{\perp}$ (mV/nm)= $135\sqrt{B_{\perp} ({\rm Tesla})-4}$  (dotted line, Fig.~\ref{Figfour} A). The non-monotonic $\sigma_{xx} (E_{\perp})$ dependence demonstrates that the first-order phase transition between the spin-unpolarized strong field state and the spin-polarized intermediate field is tuned in favor of the spin-polarized state by non-zero values of $E_{\perp}$. A conductivity peak at the transition is expected due to boundary state transport along domain walls that separate states with zero and non-zero values of the valley Hall conductivity. 
	
The multiple phases observed in the rich phase diagram arise from an intricate interplay between spin and crystalline symmetries, localization, Zeeman energy, exchange interactions and self-energies of bands with different parities. Multiband Dirac systems such as ABA-stacked multilayer provide a rich playground to realize exotic symmetry protected interacting topological phases~\cite{ref40}. Additionally, our observations open the possibility of discovering a plethora of gate tunable symmetry protected topological phases protected by point and space group symmetries in 2D crystals and heterostructures.

\acknowledgements{We thank Maxim Kharitonov for discussions. The experiments are supported by DOE BES Division under grant no. ER 46940-DE-SC0010597. The theoretical works and collaboration between theory and experiment is enabled by SHINES, which is an Energy Frontier Research Center funded by DOE BES under Award SC0012670. Part of this work was performed at NHMFL that is supported by NSF/DMR-0654118, the State of Florida, and DOE. Growth of hexagonal boron nitride crystals was supported by the Elemental Strategy Initiative conducted by the MEXT, Japan and a Grant-in-Aid for Scientific Research on Innovative Areas “Science of Atomic Layers” from JSPS.  Work in Austin was supported in part by DOE Division of Materials Sciences and Engineering grant DE-FG03- 02ER45958 and by Welch foundation grant F1473.}

%\showacknow % Display the acknowledgments section

% \pnasbreak splits and balances the columns before the references.
% If you see unexpected formatting errors, try commenting out this line
% as it can run into problems with floats and footnotes on the final page.
%\pnasbreak

% Bibliography
%\bibliography{pnas-sample}

\begin{widetext}

\section*{Supplemental Information: Quantum Parity Hall Effect in ABA Graphene}

Multi-layer graphene structures are realized by stacking 2D graphene crystals along the vertical direction. 
Stable structures of few-layer graphene can exist in a variety of different stacking sequences. 
Depending on the stacking sequence and resulting crystal symmetry, massless and massive Dirac bands can 
occur and even co-exist at or near neutral charge density~\cite{grapheneQHreview}. 

The lattice structure is important in the determination of all multilayer graphene 
band structures.   The lattice structure of ABA-trilayer graphene has the inequivalent atomic sites of the $i^{th}$ graphene layer ($A_{i}$ and $B_{i}$) stacked so that only half of the sub-lattice sites in each layer ($B_{1}, A_{2}, B_{3}$) have a near-neighbor in the adjacent layer, whereas the other half ($A_{1}, B_{2}, A_{3}$) don't have a near neighbor in the adjacent layers (see Fig. S\ref{suppfigone}). ABA-trilayer graphene is invariant under the 
$D_{3h}$ point group, which includes mirror symmetry about the middle layer. 
The sub-lattice orbital combinations $X_{\pm} = (X_{1} \pm X_{3})/\sqrt{2} $ (where $X=A,B$), form the irreducible representations of this mirror symmetry. The symmetric (anti-symmetric) combinations of the top and bottom layer orbitals $X_{\pm}$ are even (odd) with respect to this mirror symmetry, while the middle layer $X_{2}$ orbitals have even parity. Application of an external potential difference $\Delta_{1}$ between the layers breaks this mirror symmetry. 
%In this case, the even and odd parity eigenstates hybridize. 
As this work demonstrates, the electric field serves as an important experimental knob to analyze the symmetries of the 
$\nu =0$ QH state in ABA trilayer graphene.  

\section*{Non-interacting Hall conductance at $\nu=0$}

In order to understand the Landau level (LL) spectrum in ABA-trilayer graphene it is convenient to separate the 
Hamiltonian into contributions from subspaces with definite parity with respect to the mirror symmetry. 
The odd parity orbitals $(A_{-},B_{-}) $ exhibit a gapped Dirac-like dispersion similar to that of monolayer graphene (MLG), whereas the even parity orbitals $(A_{+},B_{2},A_{2},B_{+})$ exhibit a band dispersion similar to gapped bilayer 
graphene (BLG). The gaps in the odd parity MLG and even parity BLG-like bands are
due to direct interlayer remote hopping terms. 
The effective Hamiltonian~\cite{AbaninABA} for the odd parity bands in the $(A_{-}, B_{-})$ basis for valley 
${\bf K}$ and $(B_{-}, A_{-})$ for valley ${\bf K}' $ is given by,
\begin{displaymath}
\mathcal{H}_{odd}= \left( \begin{array}{cc}
\delta_{2} - \gamma_{2}/2 &  v \pi^{\dagger} \\
v \pi & -\gamma_{5}/2 + \delta + \delta_{2} \end{array} \right),
\end{displaymath}
where $\pi = \pi_{x} + i \pi_{y}$ is the momentum operator in a magnetic field, $v = 3/2 \gamma_{0} a_{0}$, where $\gamma_{0} =3100$ meV and $a_{0}$ is the C-C bond distance.
Here and below, $\gamma_{i}$ parameters denote the hopping matrix elements,
whose values are obtained by fitting our observations to crossing features in the LL spectrum. 

In the even parity sector the direct interlayer hopping $\gamma_{1} = 390$ meV pushes the even parity states $(A_{2}, B_{+})$ away from the neutral Fermi energy. The effective Hamiltonian~\cite{AbaninABA} in terms of the low-energy bands $(A_{+},B_{2})$ for valley ${\bf K}$ and $(B_{2},A_{+})$ for valley ${\bf K}'$ is,
\begin{displaymath}
\mathcal{H}_{even}= \left( \begin{array}{cc}
\gamma_{2}/2  + \delta_{2} &  -\frac{1}{2m} (\pi^{\dagger})^2 \\
-\frac{1}{2m} (\pi)^2 & -2\delta_{2} \end{array} \right)
+ \frac{v^2}{2 \gamma_{1}^2} \left( \begin{array}{cc}
(\delta - 2 \delta_{2}) \pi^{\dagger} \pi &  0 \\
0 & (\gamma_{5}/2 + \delta + \delta_{2}) \pi \pi^{\dagger} \end{array} \right).
\end{displaymath}
By fitting the LL spectrum implied by the band Hamiltonian to experiment we obtain 
$\gamma_{2} = -14.5 {\rm meV}, \gamma_{5} = 13 {\rm meV}$ and $\delta = 15$ meV. 
The parameter $\delta_{2}$ is a function of the cite energy difference between the top and bottom layers denoted $\Delta_{1}$.
For balanced layers the best fit value is $\delta_{2} =5.7$ meV. 
Fig. S\ref{suppfigone}a shows the LL band spectrum of ABA-trilayer graphene calculated with these parameter values.  

\begin{figure}[h]
\begin{center}
\includegraphics[width=0.8\textwidth]{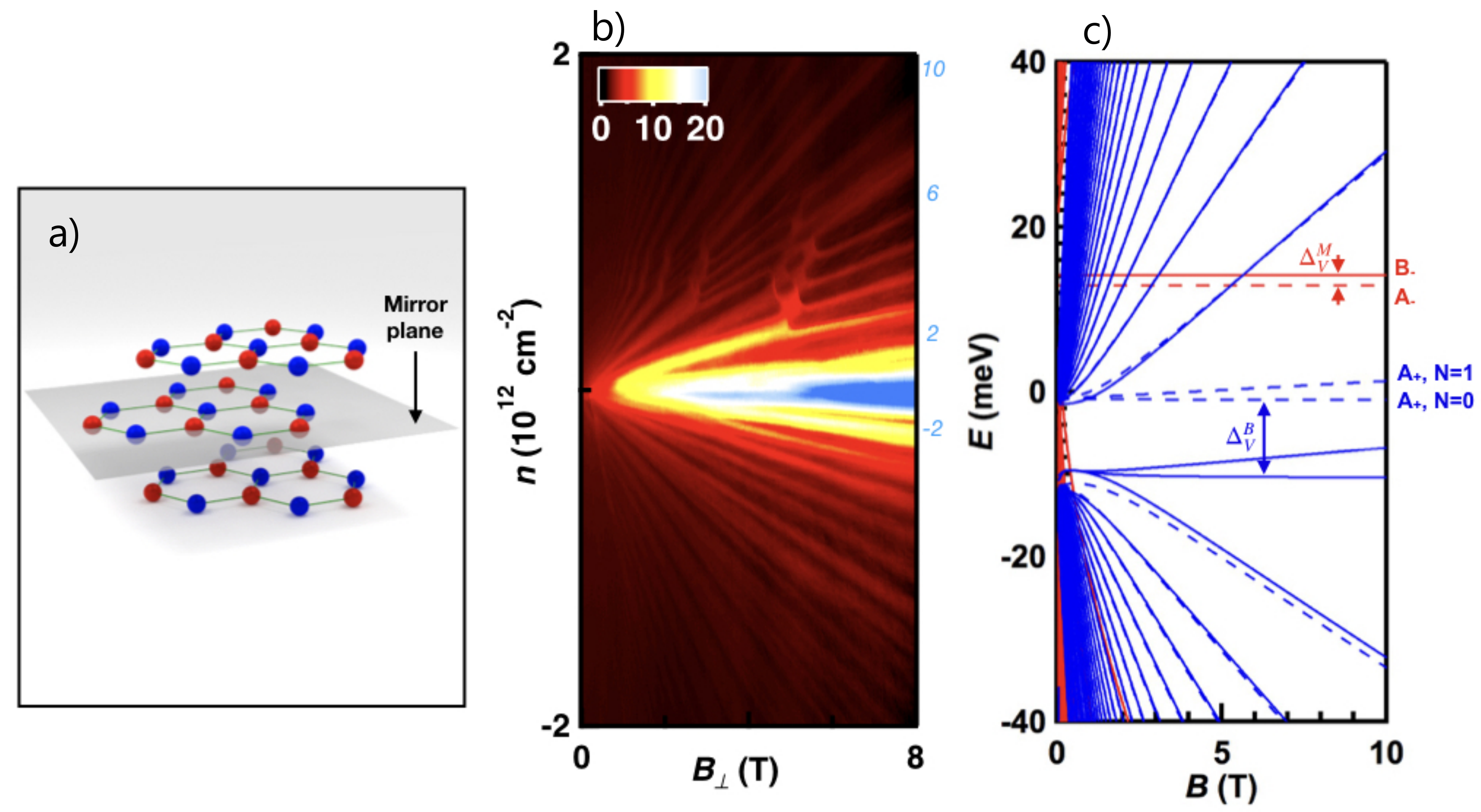}
\caption{(a) Crystal structure of ABA-trilayer graphene in which the middle layer is a mirror plane. 
(b) Landau fan diagram for ABA-trilayer graphene. (b). Full parameter k.p model calculations of the LL energy spectrum as a function of the B field. The dashed (solid) lines are for K (-K) valley Landau level states and the 
red and blue lines are for LL states arising from MLG bands that are odd under 
mirror reflection, and BLG-like bands that are 
even under mirror reflection, respectively.  Note that the $N=0$ LL energies 
are more weakly magnetic-field dependent and more strongly valley-dependent.}
\label{suppfigone}
\end{center}
\end{figure}

For balanced layers, the total filling factor $\nu = \nu_{e} + \nu_{o}$ can be written as the sum of the local BLG-like even parity band filling factor $\nu_{e}$ and the MLG-like odd parity band filling factor $\nu_{o}$.  
Important features of the LL spectrum of ABA trilayer graphene allow for the realization of the quantum parity Hall states at $\nu =0$.
First all LL energies except for those of the $N=0$ states are strongly magnetic field dependent and the
$N=0$ levels have respectively four and eight flavor components for MLG and BLG. 
Most importantly, the odd N=0 level lies above the even $N=0$ level, and the valley splitting is 
larger for the even $N=0$ level than for the odd $N=0$ level.
In a single-particle theory, these properties conspire with the neutrality condition
$ \nu_{e} + \nu_{o} =0$ to imply that above a very weak critical magnetic field
$\nu_{o}=-2$ ( empty odd $N=0$ levels ) and $\nu_{e}= +2$ (3/4 filled 
even $N=0$ levels).  
 The eigenstates and energy eigenvalues of the odd parity orbitals of the zeroth LL in the Landau gauge (${\bf A} = (0,-Bx,0) $) is given by,
\begin{displaymath}
\Phi_{0,{\bf K}} = \left( \begin{array}{c} 0 \\
\phi_{0,k_y}(\vec{x}) \end{array} \right) ; E = \delta - \gamma_{5}/2 + \delta_{2}, \qquad 
\Phi_{0,{\bf K'}} = \left( \begin{array}{c} 0 \\
\phi_{0,k_y}(\vec{x}) \end{array} \right); E = - \gamma_{2}/2 + \delta_{2},
\end{displaymath}
where $\phi_{n,k_y}(\vec{x}) = \langle \vec{x} | n,k_{y} \rangle $ is the real space $n^{th}$ LL wavefunction and each eigenstate is spin degenerate. For the zeroth LL $A_{-}$ orbitals occupy the ${\bf K}'$ valley and $B_{-}$ orbitals
the ${\bf K}$ valley. The $A_{-}$ orbital is at a lower energy than the $B_{-}$ orbital with a 
valley or orbital splitting $\Delta^{M}_{V} = 1.25$ meV induced by the interlayer hopping terms, and
when mirror symmetry is broken by an external displacement field also the potential 
energy difference between the layers. 

The even parity eigenstates and corresponding energy eigenvalues for the zeroth LL are:  
\begin{displaymath}
\Phi_{0,K} = \left( \begin{array}{c} 0 \\
\phi_{0,k_y} (\vec{x}) \end{array} \right); E = -2 \delta_{2}
\quad 
\Phi_{1,K} = \left( \begin{array}{c} 0 \\
\phi_{1,k_y}(\vec{x}) \end{array} \right); E = -2\delta_{2} + \frac{h^2 v^2}{\gamma_{1}^2 l^2_{B}} (\delta_{2} + \delta + \gamma_{5}/2),
\end{displaymath}
which reside completely on sublattice $B_{2}$, and 
\begin{displaymath}
\Phi_{0,K'} =\left( \begin{array}{c} 0 \\
\phi_{0,k_y}(\vec{x}) \end{array} \right); E = \gamma_{2}/2  
\quad 
\Phi_{1,K'} = \left( \begin{array}{c} 0 \\
\phi_{1,k_y} \vec{x} \end{array} \right); E = \gamma_{2}/2 + \frac{h^2 v^2}{\gamma_{1}^2 l^2_{B}} (\delta - 2\delta_{2}),
\end{displaymath}
which reside on sub-lattice $A_{+}$. 
As before remote interlayer hopping and the energy difference between the layers determines
the valley splitting in the bilayer.  In the absence of a gate field
$\Delta^{B}_{V}= |7.25 \, {\rm meV} - 3 \delta_{2}| = 9.85 {\rm meV}$. Since $\gamma_2 > -6 \delta_2$, 
the holes in the even parity zeroth LL reside on sub-lattice $B_2$, while the electrons in the even parity zeroth LL 
reside on sub-lattice $A_{+}$.

As illustrated in Fig. S\ref{suppfigone}a), the conduction-band portion of the 
BLG-like $N=0$ Landau level is half-occupied at neutrality, whereas the MBG-like $N=0$ 
Landau level is empty.  This implies that the spin degenerate gapless edge modes of the MLG-like hole branch and 
the BLG-like electron branch flow in opposite directions.  However 
the counter-propagating MLG-like hole branch and the BLG-like electron branch don't interact as long the 
mirror symmetry is preserved. This mirror symmetry
prohibits back-scattering events, leading to a robust plateau in the two-terminal conductivity
with $\sigma_{xx} =4e^2/h$, consistent with the experimental observations. 
Because it breaks mirror symmetry, a displacement filed that acts between the layers mixes even and odd subspaces.
In the Landau gauge (${\bf A} = (0,-Bx,0) $), its projection onto the $N=0$ subspace is 
\begin{equation}
H_{\Delta} = \frac{\Delta_{1}}{2} \sum_{\sigma} (c^{\dagger}_{A_{+},\sigma,X}  c_{A_{-},\sigma,X} +c^{\dagger}_{A_{-},\sigma,X}  c_{A_{+},\sigma,X} ),
\end{equation}
where $\Delta_{1} (E)$ scales linearly with the displacement field.
This terms in the Hamiltonian will introduce a gap in the edge state spectrum and 
allow back scattering due to disorder.  
This explains the rapid reduction in the longitudinal conductivity from $ 4e^2/h \to 0$
when a displacement field is applied. 

To understand the rich phase diagram observed at higher $B_{\perp}$ we must look more closely at the 
orbital character of the various LL states.  
The $N=0$ odd-parity band Hamiltonian projected onto the zeroth LL is,
\begin{equation}
H_{odd} = \sum_{X,\tau, \sigma} \bigg[\frac{\Delta_{mb}}{2} + \frac{\Delta^{M}_{V}}{2} (\mathcal{I} \otimes \hat{\tau}_{z} ) + \frac{\Delta_{z}}{2} (\hat{s}^{z} \otimes \mathcal{I}) \bigg] c^{\dagger}_{\tau,\sigma,X}  c_{\tau',\sigma',X},
\end{equation}
where we have made a convenient choice for the zero of energy,
the Pauli matrices $\hat{\tau}_{i}$ and $\hat{s}_{i}$ act on the orbital/valley and spin degrees of freedom respectively, $\tau,\sigma$ are valley and spin labels, $\Delta_{Z} $ is the Zeeman energy and $\Delta_{mb} = 11.5 - 3 \delta_2/2 =20.05 {\rm meV} $ is the even/odd parity energy splitting. Similarly for the BLG-like even parity bands we can write the zeroth LL projected band Hamiltonian as,
\begin{equation}
H_{even} = \sum_{X,\tau,\sigma,\alpha} \bigg[-\frac{\Delta_{mb}}{2} + \frac{\Delta^{B}_{V}}{2} (\mathcal{I} \otimes \hat{\tau}_{z} ) + \frac{\Delta_{z}}{2} (\hat{s}^{z} \otimes \mathcal{I}) + \Delta^{\tau}_{LL} \delta_{\alpha,1}\bigg] c^{\dagger}_{\alpha,\tau,\sigma,X}  c_{\alpha',\tau,\sigma,X}, 
\end{equation}
where $\alpha,\alpha' =0,1$ denote the LL pseudospin orbital index in the zeroth LL, $\Delta^{\tau}_{LL}$ is the orbital dependent LL pseudo-spin gap and $\delta_{\alpha,1}$ is the Kronecker delta symbol. 

\section*{Mirror symmetry in ABA-trilayer graphene}

The lattice structure of ABA-trilayer graphene has a mirror symmetry with respect to the middle layer. Since ABA-trilayer is mirror symmetric the Hamiltonian $\mathcal{H} = H_{even} \otimes  H_{odd}$ must commute with the mirror symmetry operator $\hat{\mathcal{M}}$. In the case of a high magnetic field, this implies that the Landau level wavefunctions can be simultaneously labeled by the eigenvalues of $\mathcal{H}$ and $\hat{\mathcal{M}}$. Below, we describe the mirror symmetry operator for ABA trilayer graphene for spinless particles and particles with spin.

\subsection*{Spinless case}

In general, the mirror symmetry operator $\hat{\mathcal{M}}$ can be expressed as a rotation by $\pi$ with the axis of rotation perpendicular to the mirror plane, followed by an inversion. This gives $\hat{\mathcal{M}}= P D(\pi)$, where $P$ is the inversion operator which send $\vec{r} \to - \vec{r}$, and $D(\pi)$ is the rotation operator which acts on the internal degrees of freedom such as spin. For spinless particles, this rotation matrix acts trivially and is given by the identity matrix. Under mirror symmetry the orbital in the top layer are interchanged with the orbitals in the bottom layer $\psi_{A_{1}} \leftrightarrow \psi_{A_{3}}$ and $\psi_{B_{1}} \leftrightarrow \psi_{B_{3}}$ while the orbitals in the middle layer remain unchanged $\psi_{A_{2}} \leftrightarrow \psi_{A_{2}}$ and $\psi_{B_{2}} \leftrightarrow \psi_{B_{2}}$. The mirror symmetry operator in the orbital basis can be expressed as
\begin{displaymath}
\hat{\mathcal{M}} = \left( \begin{array}{cccccc} 
0 & 0 & 0 & 0 & 1 & 0 \\
0 & 0 & 0 & 0 & 0 & 1 \\
0 & 0 & 1 & 0 & 0 & 0 \\
0 & 0 & 0 & 1 & 0 & 0 \\
1 & 0 & 0 & 0 & 0 & 0 \\
0 & 1 & 0 & 0 & 0 & 0
\end{array} \right) ,
\end{displaymath}
which acts on the orbital basis $\Psi=(\psi_{A_1},\psi_{B_1},\psi_{A_2},\psi_{B_2},\psi_{A_3},\psi_{B_3})$. In this case $\hat{\mathcal{M}}^2=1$ and the $\hat{\mathcal{M}}$ has the eigenvalues $\pm 1$.

\subsection*{Particles with spin}
When the spin degree of freedom is included the rotation matrix operator $D(\pi)$ acts non-trivially on the spinor wavefunction. In this case the rotation operator acts on the spin degree of freedom and can be expressed as, 
\begin{displaymath}
D(\pi) = \exp(-i \frac{\pi \sigma^z}{2}) = \left( \begin{array}{cc} 
-i & 0 \\ 
0 & i 
\end{array} \right) = -i \sigma_{z} .
\end{displaymath}
Combined the mirror symmetry operator for particles with spin can then be expressed as 
\begin{displaymath}
\hat{\mathcal{M}} = \left( \begin{array}{cc} 
-i\hat{\mathcal{M}} & 0 \\ 
0 & i \hat{\mathcal{M}}
\end{array} \right),
\end{displaymath}
which acts on the 12-component spinor $\Psi = (\Psi_{\uparrow}, \Psi_{\downarrow})$. In the
spin space $\psi_{A_{1} \uparrow} \leftrightarrow  -i \psi_{A_{3} \uparrow}$ and $\psi_{A_{1} \downarrow} \leftrightarrow i \psi_{A_{3} \downarrow}$ with the same relation for the $B_{1}$ and $B_{3}$ orbital. This implies that for particles with spin $\hat{\mathcal{M}}^2=-1$, is an anti-unitary operator. Therefore, in this case the eigenstates of different parities can be labeled by the eigenvalues $\pm i $. In particular, the eigenvalues of the $\hat{\mathcal{M}}$ for the even parity eigenstates are given by $+i(-1)^{\sigma}$, while the odd parity eigenstates are $-i(-1)^{\sigma}$, where $\sigma=\pm$ for the $\uparrow(\downarrow)$ spins. This implies the projected mirror symmetry operator $\overline{\mathcal{M}} $ for the spin polarized state obeys $\overline{\mathcal{M}}^2 =1$. This provides an effective Kramer degeneracy for spin states in each mirror sector. 

As the magnetic field is increased electron-electron interactions within the zeroth LL significantly modify the $\nu =0$ QH state. Interactions can lift the spin degeneracy of the $\nu=0 $ QH state while at the same time lowering the relative zero-point energy of the electron and hole LLs resulting in a QH ferromagnetic state with counter-propagating edge states with opposite spins. This results in a quantized two-terminal conductance $\sigma_{xx} = 2 e^2/h$ which is protected by mirror symmetry. We discuss the effect of Coulomb interactions in the $\nu =0$ QH state of ABA-trilayer graphene in the next section.

\section*{Interaction energy in ABA-trilayer graphene}   
     
It is well known that electron-electron interactions in graphene and semiconducting 2DEGs result in interaction induced QH plateaus not predicted by Landau quantization alone. In multi-component systems, interaction induced incompressibilities  normally result from broken symmetries which lower ground state energies.
% The Coulomb interaction energy $V_{int}\sim \sqrt{B}$, becomes stronger at high magnetic fields. 
In ABA-trilayer graphene these interaction induced broken symmetry ordered states compete with band 
gaps induced by Landau quantization. 
%This sequence of the zeroth LL determined primarily by the remote hopping parameters of the $k \cdot p$ model Hamiltonian. 
In this section, we express the mean field Hartree-Fock energy
as a function of an order parameter characterizing all possible broken symmetry states
in the 12-fold degenerate $N=0$ space of ABA-trilayer graphene.
        
Assuming sublattice and valley independent Coulomb interactions, the interaction Hamiltonian can be expressed as,
\begin{equation} 
\label{ham}
\mathcal{H}_{int} = \frac{1}{2 L^2} \sum_{{\bf q}} v_{{\bf q}} \rho_{{\bf q}} \rho_{-{\bf q}} ,
\end{equation}
where $v_{{\bf q}} = 2\pi e^2/(\epsilon q)$ and $\rho_{{\bf q}} = \sum_{k} \psi_{{\bf k}+ {\bf q}, X_{i}, \sigma}\psi_{{\bf k}, X_{i}, \sigma} $, $X_{i}$ denotes the sublattice degree of freedom for the $i^{th}$ layer and $\sigma$ is the spin and valley index. In the zeroth LL of ABA-trilayer graphene the valley and layer/sub-lattice degrees of freedom are equivalent. 
For balanced ABA-trilayer ($\Delta_{1}=0$) the electron density can be expressed in terms of fields with well-defined parity $\rho_{{\bf q}} =\rho^{e}_{{\bf q}} + \rho^o_{{\bf q}} $, where the $\rho^{i}_{{\bf q}}$ is the density field with even (odd) parity. In terms of even and odd parity fields the interaction Hamiltonian becomes,
\begin{equation}
\mathcal{H}_{int} = \frac{1}{2 L^2} \sum_{{\bf q}} v_{q} (:\rho^e_{{\bf q}} \rho^e_{-{\bf q}}: + :\rho^o_{{\bf q}} \rho^o_{-{\bf q}}:).
\end{equation}
We neglect layer interaction anisotropy; this assumption is well justified since the ratio $d/l_{B} \leq 0.1 $
where $d$ is the layer spacing and $l_{B}$ is the magnetic length.

To understand the role of interactions on the ordering of the broken symmetry states, we must express the Coulomb interaction in terms of the LL spectrum of ABA-trilayer graphene. For the even parity states the sublattice orbitals ($B_{+}$ and $A_{2}$) are pushed away from the Fermi energy by $\sim \gamma_{1}$. 
Therefore, at low-energies we only need the sublattice orbitals ($A_{+} $ and $ B_{2}$). 
The Coulomb interaction projected on the zeroth LL becomes, 
\begin{equation}
\mathcal{H}_{int} = \frac{1}{2 L^2} \sum_{{\bf q}} v_{{\bf q}} (:\bar{\rho}^e_{{\bf q}} \bar{\rho}^e_{-{\bf q}}: + :\bar{\rho}^o_{{\bf q}} \bar{\rho}^o_{-{\bf q}}:), 
\end{equation}
where the projected even(odd) density operator in the Landau gauge is defined as,
\begin{eqnarray}
\bar{\rho}^{e}_{{\bf q}} &=& \sum_{n,n',\sigma} \sum_{X,X'} F_{n,n'}({\bf q}) \delta(q_y l_{B}^{2} + X -X') e^{\frac{-iq_{x}}{2}(X +X')} b^{\dagger}_{n,k_y,\sigma} b_{n',{k'}_y,\sigma}, \\
\bar{\rho}^{o}_{{\bf q}} &=& \sum_{\sigma} \sum_{X,X'} F_{0,0}({\bf q}) \delta(q_y l_{B}^{2} + X -X') e^{\frac{-iq_{x}}{2}(X +X')} a^{\dagger}_{n,k_y,\sigma} a_{n',{k'}_y,\sigma},  
\end{eqnarray}
where $n,n'=0,1$ denote the extra LL index always present in BLG $N=0$ Landau levels, 
$\sigma$ denotes spin and valley $=$sublattice index, and $X, X'$ are the guiding center quantum numbers. 
The $b^{\dagger}/a^{\dagger}$ are the creation operators for the even and odd parity LLs respectively. 
The form factors $F_{n,n'}({\bf q})$  (with $F_{00} ({\bf q}) = e^{-(ql_{B})^{2}/4}$, $F_{10} ({\bf q}) = (i q_{x}+q_{y}) l_{B)} e^{-(ql_{B})^{2}/4}/\sqrt{2} = [{F}_{01}(-\bf{q})]^*$ and $ F_{11}({\bf q}) = (1-(ql_{B})^{2}/2)e^{(-ql_{B})^{2}/4}$), reflect the character of the two different quantum cyclotron orbits in the zeroth LL. Self-energy corrections will be included to account for LL mixing justifying our projection onto the zeroth LL.
 
At neutrality, six of the twelve $N=0$ Landau level orbitals are occupied.
To calculate the Hartree-Fock mean field energy we therefore make the following generalized quantum Hall ferromagnetic wave function {\it ansatz} for the translationally invariant ground state at $\nu = 0$,
\begin{equation}
| \Psi_{0} \rangle = \prod_{X}  \bigg( \prod_{i=1}^{6}  z^{\alpha_{i}}_{\sigma_{i}} c^{\dagger}_{\sigma_{i},X} \bigg) |\Omega \rangle,
\end{equation}
where the Einstein summation convention applies to $\sigma_{i}$'s and the six occupied 
12-component spinors must be orthogonal. 
$|\Omega \rangle $ is the vacuum state in which all hole-like $N \neq 0$ Landau 
levels are occupied for both the even and odd parity orbitals. 
The partial filling of the zeroth LL is represented by an order parameter that captures the possible ordering of the broken symmetry state. The order parameter can be written as $\Delta= \sum_{\alpha} z^{\alpha} \bar{z}^{\alpha}$. To satisfy charge neutrality we impose $Tr[\Delta] = 6 $ and $Tr[\Delta^2]=6$. 
It is convenient to express the order parameter in terms of nine $4 \times 4$ matrices as
\begin{displaymath}
\Delta = \left( \begin{array}{ccc}
\Delta^{e}_{00} &  \Delta^{e}_{01} & \Delta^{eo}_{0} \\
\Delta^{e}_{10} &  \Delta^{e}_{11} & \Delta^{eo}_{1} \\
\Delta^{oe}_{0} & \Delta^{oe}_{1} & \Delta^{o}\end{array} \right).
\end{displaymath}

The Hartree-Fock energy of the $\nu =0 $ ground state is $E_{int} =\langle \Psi_{0}|\mathcal{H}_{int}| \Psi_{0} \rangle$. The mean field energy based on the wavefunction ansatz becomes,
\begin{equation}
\mathcal{E}_{int} = \frac{E_{int}}{N_{\phi}} = \frac{1}{2} \sum_{\{n\}_{i}; \sigma', \sigma} H_{n_{1},n_{2},n_{3},n_{4}} Tr[\Delta^{e}_{n_{1} n_{4}}] (Tr[\Delta^{e}_{n_{1},n_{4}}])^2 -  \frac{1}{2} X_{n_{1},n_{2},n_{3},n_{4}} Tr[\Delta^{e}_{n_{1} n_{3}} \Delta^{e}_{n_{2} n_{4}}],
\end{equation}
where $n_{i} $ denotes the LL band index of the four-fold degenerate LLs of the odd parity bands and we define,
\begin{equation}
H_{n_{1},n_{2},n_{3},n_{4}} = \lim_{q \to 0} v_{q} F_{n_{1},n_{4}} (q) F_{n_{2},n_{3}} (-q),
\end{equation}
which captures the electrostatic interaction. This direct energy can be absorbed by the background and doesn't effect the broken symmetry ordering for $d/l_{B} \sim 0$.
\begin{equation}
X_{n_{1},n_{2},n_{3},n_{4}} = \frac{1}{L^2} \sum_{q} v_{q} F_{n_{1},n_{4}} (q) F_{n_{2},n_{3}} (-q),
\end{equation}
is the projected exchange interaction that is essential for the spontaneous ordering of the spin and valley quantum degrees of freedom. Neglecting the Hartree term, the interaction energy contribution from zeroth LL odd parity orbitals is
\begin{equation}
\mathcal{E}^o_{int} = \frac{E^{o}_{HF}}{N_{\phi}} = - \frac{1}{2} \bigg( X Tr[\Delta^{o} \Delta^{o}] - Tr [\Sigma^{o} \Delta^{o}] \bigg),
\end{equation}  
where $X = \sqrt{\pi/2} [e^2/(\epsilon l_{B})]$ is the interaction exchange energy. $\Sigma$ denotes the self-energy which captures the interaction of the partially filled zeroth LL with the sea of negative energy LLs. The self-energy can be defined as 
\begin{equation}
\Sigma_{\sigma,\sigma'} = -\frac{1}{2} \sum_{n_{1},n_{2}} X_{n_{1},0,n_{2},0} [\Delta^{e}_{n_{1},n_{2}} ]_{\sigma,\sigma'},
\end{equation}
where $\sigma, \sigma'$ denote the four-fold spin and valley degeneracy. The interaction of the partially filled zeroth LL orbitals with the negative energy LLs results in self-energy corrections to the zeroth LLs. In single-band systems, the self-energy of the LL orbitals can be neglected by renormalization of the zero point energy. However, in multi-band systems, such as ABA trilayers, the relative self-energies can affect the ordering of the partially filled LLs. 
This self-energy interaction is essential for understanding the various transitions in the $\nu =0$ state of ABA-trilayer graphene.

The mean field energy contribution from 
even parity orbitals can be expressed in a similar way:
\begin{eqnarray}
\mathcal{E}^{e}_{int} &=& -\frac{1}{2} X_{0000} Tr[\Delta^e_{00} \Delta^e_{00}] -\frac{1}{2} X_{1111} Tr[\Delta^e_{11} \Delta^e_{11}]- X_{0110} Tr[\Delta^e_{00} \Delta^e_{11}] \\ \nonumber
&-&  X_{0011} Tr[\Delta^e_{01} \Delta^e_{01}] + Tr[\Sigma^{e}_{0} \Delta^e_{00}] +Tr[\Sigma^{e}_{1} \Delta^{e}_{11}],
\end{eqnarray}
where the zeroth LL of the even parity orbitals contains the LL pseudospin orbital degree of freedom $n=0,1$. 
We account for different self energies for the LL-pseudospin orbital $\Sigma^{e}_{\alpha}$ where $\alpha =0,1$ denotes the LL pseudospin orbital. The total mean field energy becomes 
\begin{eqnarray}
\mathcal{E}_{int} &=& -\sqrt{\frac{\pi}{2}} \frac{e^2}{\epsilon l_{B}} \bigg[ \frac{1}{2} Tr[\Delta^e_{00} \Delta^e_{00}] + \frac{3}{8} Tr[\Delta^e_{11} \Delta^e_{11}] + \frac{1}{2} Tr[\Delta^e_{00} \Delta^e_{11}] \\ \nonumber
&+& \frac{1}{2} Tr[\Delta^e_{01} \Delta^e_{01}] + \frac{1}{2} Tr[\Delta^o \Delta^o] \bigg] + Tr[\Sigma^{o} \Delta^o] 
+ Tr[\Sigma^{e}_{0} \Delta^e_{00}] +Tr[\Sigma^{e}_{1} \Delta^{e}_{11}] .
\label{eq:eint} 
\end{eqnarray}

The final three terms in Eq.~\ref{eq:eint} account for exchange interactions between $N=0$ Landau level 
states and the negative energy sea.  Because the the bare band parameters implicitly include 
interactions with a fully occupied set of valence band states for both MLG and BLG, this 
self-energy is particle-hole symmetric and positive for positive filling factors.
The increase in self-energy
with filling factor is the strong magnetic field counterpart of the self-energy effects 
that are responsible for the interaction enhancement of the velocity at the Dirac point in 
monolayer graphene. 
Its role is to lower the energy of the hole-like LLs and raise the energy of the electron-like LLs
by an amount that increases monotonically with the strength of
the perpendicular magnetic field $B_{\perp}$. 
In addition to lifting the spin, valley and LL pseudo-spin degeneracy of the zeroth LLs,
interactions with therefore also tends to favor states with smaller $\nu_{e} = |\nu_{o}|$.  
This competition between interaction energies within the $N=0$ manifold,
the self-energy due to interactions with the negative energy sea, and the single band LL structure results in 
a rich phase diagram which we now relate to the experimental observations for 
neutral charge density $\nu = \nu_{e}+\nu_{o} =0$ in ABA-trilayer graphene.  

\section*{Phase diagram of the $\nu=0$ QH state in ABA trilayer graphene}

The experimental phase diagram of the $\nu = 0$ QH state in ABA-trilayer graphene is extremely rich. 
%Mirror symmetry, valid only for balanced ABA trilayer graphene ($\Delta_{1} =0$), implies that the filling factor $\nu$ can be expressed as $\nu = \nu_{o} + \nu_{e}$, where $\nu_{e}$ represents the partial occupation of the electron LLs of the even parity (BLG-like) bands and $\nu_{o}$ represents the partial occupation of the hole-LLs of the odd parity (MLG-like) bands. At $\nu =0$ which corresponds to partial filling of the zeroth LL, half of the degeneracies are occupied. While the Hall conductivity should vanish in all these case the diagonal conductivity is sensitive to the local filling even and odd parity band of the zeroth LL.   
For balanced layers, the $\nu=0$ state exhibits transition between three phases, and 
additional phases appear when external in-plane magnetic and displacement fields are applied.
These transitions result from the magnetic field dependence of the interaction strength which 
yields a typical energy scale  $e^2/(\epsilon l_{B}) \sim (56.1/\epsilon) \sqrt{B}$. 
For balanced layers interaction effects compete with the Landau quantization sequence determined of the band Hamiltonian.  Below, we describe the symmetries of the interacting broken symmetry states and calculate their energies. 

\subsection{Quantum Parity Hall State}

This first phase, which appears at low magnetic fields, exhibits a quantized value of the
longitudinal conductance $\sigma_{xx} = 4e^2/h$. 
This phase is stabilized mainly by single-particle physics 
and its transport features can be understood from the LL bands of ABA-trilayer graphene.  
Band structure calculations show that some electron-like even parity
$N=0$ LL states are lower in energy than  the hole-like odd-party $N=0$ LL states.
%The co-presence of the spin degenerate electron and hole-like LLs belonging to different representations of the mirror symmetry result to counter-propagating edge modes indicated by the quantization in the longitudinal conductivity. 
Because scattering between even-parity and odd-parity states,
due to either band or disorder effects, is prohibited by 
mirror symmetry, this state has spin degenerate counter-propagating edge modes resulting in 
quantization of the longitudinal conductance $\sigma_{xx} = 4e^2/h$.  
This state is a symmetry protected topological state and exhibits quantum parity Hall effect.  
When a displacement field is applied the mirror symmetry 
of the lattice is broken and the quantization of the diagonal conductance is destroyed. 
The sharp decline of the diagonal conductivity results from back-scattering events that lead to hybridization of the 
even and the odd parity bands. We call this state the quantum parity Hall (QPH) state. 

The total filling factor is given by $\nu = \nu_{e} + \nu_{o} = 2 +(-2)$. The hole LLs have the quantum numbers$(A_{-},\uparrow)$ and $(A_{-},\downarrow)$. However, there are two possibilities for the electron LLs $(A_{+},\uparrow, 0)$ and $(A_{+},\uparrow, 1)$, which is spin polarized, or $(A_{+},\uparrow, 0)$ and $(A_{+},\downarrow, 0)$ which is LL pseudo-spin polarized. The single particle gaps $\Delta_{LL} > \Delta_{z}$ and $\Delta^{M}_{V} > \Delta_{z}$ dictate that the electron-like BLG LLs have the quantum numbers $(A_{+},\uparrow, 0)$ and $(A_{+},\downarrow, 0)$, while unoccupied the hole-like MLG LLs contain the quantum numbers $(A_{-},\uparrow)$ and $(A_{-},\downarrow)$. The interaction energy for the QPH state is 
\begin{equation}
\mathcal{E}_{QPH} = -\frac{15}{4} \sqrt{\frac{\pi}{2}} \frac{e^2}{(\epsilon l_{B})} - \Delta^{B}_{V} - 2 \Delta_{LL} - 3 \Delta_{mb} - 3 \Sigma^e_{1} 
\end{equation}
We would like to point out that interaction effects in the even parity branch can cause a transition from the LL pseudo-spin polarized state to a spin polarized state.
This state would also exhibit a quantized longitudinal conductivity $\sigma_{xx} = 4e^2/h$. However, there is no experimental evidence of this transition. As the magnetic field is increased above $B_{c1}$ ($B_{c1} \sim 1$T) there is a transition in the longitudinal conductivity due to interaction effects, which we discuss next.

\subsection{Quantum Parity Hall Ferromagnet}

At higher fields the longitudinal conductivity is reduced to half of its non-interacting value $\sigma_{xx}= 2e^2/h$. 
This occurs as the combined effect of interactions within the $N=0$ LLs and exchange interactions with 
the negative energy sea.  
%Self-energy corrections due to interactions with the Dirac sea of negative energy LLs lowers the energy of the hole-like LLs while raising the energy of the electron-like LLs. The intra-LL exchange interaction lifts the spin degeneracy of both the electron and hole-like LLs.
%The combined effect leads to a reduction in the longitudinal conductivity from $4e^2/h \to 2e^2/h$.
%
As explained above, the standard regularization in which the self-energy is 
referenced to its value at charge neutrality points causes 
electron-like LLs to be raised in energy relative to hole-like LLs.  
If this were the only effect of interactions in the system, we expect it to drive a transition to an insulating state 
in which both $\nu_{e}$ and $\nu_{o}$ are zero and both diagonal and Hall conductances 
vanish.  However the possibility of breaking spin degeneracy leads first to a state in 
which $\nu_{e}$ and $|\nu_{o}|$ are first reduced from two to one, and the bulk state is 
spontaneously spin-polarized.  Because $\nu_{e}$ is positive its edge state has majority spin character,
whereas $\nu_{o}$ has a counter-propagating minority spin edge channel because its filling factor is
negative.  This state is actually a spin Hall state since it has bulk Hall conductivities with opposite sign
for opposite spins. Its longitudinal conductance is reduced from $\sigma_{xx} = 4e^2/h$ in the 
quantum parity Hall state to $\sigma_{xx} = 2 e^2/h$.  The robust quantization in the longitudinal conductivity is still protected by the mirror symmetry of the crystal lattice.   As before a displacement field 
destroys this state because it enables hybridization and localization of counter-propagating edge states.  
We call this state the quantum parity Hall ferromagnet (QPHF).

The QPHF state has $\nu= 1 + (-1) $, one occupied electron-like BLG LL,
and one occupied hole-like MLG LL.  
Counter-propagating even parity spin-up electrons and odd parity spin-down holes 
holes at the edge lead to vanishing Hall conductance and a diagonal conductivity 
that is twice the conductance quantum.
The occupied electron-like BLG LL has the quantum numbers $(A_{+},\uparrow)$, while the occupied 
hole-like MLG LL the quantum numbers $(A_{-},\downarrow)$. The ground state energy is,  
\begin{equation}
\mathcal{E}_{QPHF} = -\frac{15}{4} \sqrt{\frac{\pi}{2}} \frac{e^2}{(\epsilon l_{B})} - \frac{3}{2}\Delta^{B}_{V} - 2 \Delta_{LL} - 2 \Delta_{mb} 
-\frac{\Delta^M_V}{2} - \Delta_{Z} - \Sigma^{e}_{0} - 2 \Sigma^e_{1} - \Sigma^{o}.
\end{equation}
%{\bf Allan:  Let's be careful here.  Doesn't back scattering also require spin-mixing?}  
As mentioned above, this quantum spin Hall-like state has counter-propagating even and odd parity 
edge state branches with opposite spin quantum numbers.
However, unlike the  quantum spin Hall state, in this state the quantization of the longitudinal conductivity is 
due to the mirror symmetry and not time reversal symmetry which applies to the former. 
As the mirror symmetry is broken the counter-propagating spin intermix resulting in the destruction of the plateau in the 
longitudinal conductivity. The spins intermix due to the presence of a Rashba spin orbit interaction, which is allowed as the 
displacement field in increased. Because the edges states are gapless a very small Rashba spin-orbit coupling leads to backscattering of the counterpropagating spin and hence localization of the edge states.

By comparing to experiment, the mean field energy expression can be used to determine the 
strength of exchange interactions with the negative energy sea. 
In order for the QPHF state to be lowest energy in energy 
the condition $\mathcal{E}_{QPHF} \leq \mathcal{E}_{QPH}$ must be satisfied.  This 
implies that 
\begin{equation}
\Sigma^{e}_{0} + \Sigma^{o} \geq  \Delta_{mb} -\frac{\Delta^{M}_{V} + \Delta^{B}_{V}}{2} - \Delta_{z} = 14.5 {\rm meV} - 
\Delta_{Z}.
\end{equation}
This inequality is satisfied for the $B \geq B_{c1}$ (where $B_{c1}$ is the critical magnetic field for the QPH $\to$ QPHF  transition). The right-hand side only involves the remote hopping band parameters and the Zeeman energy. Since the exchange energy is same for both 
broken symmetry states it cancels out. The above condition shows that this transition can be tuned by the magnetic field. Since the self-energy monotonically increases with $B_{\perp}$, we expect this to occur as the magnetic field is increased.

\subsection{Quantum Valley Ferromagnetic Insulator}   

As the perpendicular magnetic field is increased the longitudinal conductance 
eventually transitions from $2e^2/h \to 0$. For $B_{\perp} > 6 $T, ABA-trilayer graphene is a quantum valley ferromagnetic  insulator (QVFI). In this state, the odd parity hole-like LLs move below the Fermi energy and the even parity electron-like LLs move above the charge neutrality point. When this happens there is a clear gap at the Fermi energy which simply gives an insulator. Since there are no edge states at the Fermi level, $\sigma_{xx} \to 0$. 
The even and odd parity occupied LLs are both valley polarized and there is therefore a bulk
valley Hall effect.  

In this state, all the hole-like LLs for both even and odd parity are occupied whereas the electron-like LLs are unoccupied. The occupied even parity LLs have the quantum numbers $(B_2,n,\sigma)$, for both $n=0,1$ and $\sigma = \uparrow, \downarrow$, while the odd parity LL have the quantum numbers $(A_{-},\sigma)$. The ground state energy of the QVFI is 
\begin{equation}
\mathcal{E}_{QVFI} = -\frac{15}{4} \sqrt{\frac{\pi}{2}} \frac{e^2}{(\epsilon l_{B})}- 2\Delta^{B}_{V} - 2 \Delta_{LL} - \Delta_{mb} -\Delta^M_V - 2 \Sigma^{e}_{0} - 2 \Sigma^e_{1} - 2\Sigma^{o}
\end{equation}
Again comparison of the ground state energies $\mathcal{E}_{QPHF} \geq \mathcal{E}_{QVFI}$ gives a condition on the self-energies
\begin{equation}
\Sigma^{e}_{0} + \Sigma^{o} \geq \Delta_{mb} + \Delta_{Z} -\frac{\Delta^{M}_{V} + \Delta^{B}_{V}}{2} = 14.5 {\rm meV} + \Delta_{z},
\end{equation}
with the equality satisfied for $B = B_{c2}$ ($B_{c2}$ is the critical field for the transition from the QPHF state to the QVFI. Again, the self-energy conditions for both transitions depend on the Zeeman energy and the band parameters which are independent of the magnetic field. Therefore, the Zeeman energy causes the transition from the QPHF state to the QVFI. 
Since the Zeeman energy is small the combined inequalities also indicate that the sum of the self energies is weakly dependent on $B_{\perp}$. The condition derived above also suggests that this transition can be reversed by fixing $B_{\perp}$ in the insulating region and varying the total magnetic field. This was done by adding an in-plane magnetic field, we discuss this in the next.  

\subsection{Effect of in plane field in the QVF insulator}

The spin-polarized QPHF state is stabilized relative to the QVFI state by the Zeeman energy which 
is proportional to the total magnetic field $B=\sqrt{B_{\perp}^2+ B_{||}^2}$.  
Since the interaction self-energies are dependent on $B_{\perp}$ and other band parameters
are magnetic-field independent, the region of stability of the QPHF state can be 
enhanced by increasing the in-plane magnetic field. As before, for balanced layers ($\Delta_{1} =0$), mirror symmetry still protects the counter-propagating states. However, the spins of the counter-propagating edge states is now canted towards the direction of the total magnetic
field. 

%\begin{figure}[h]
%\begin{center}
%\includegraphics[height=3.0in]{supplementalfig2}
%\caption{Variation of the diagonal conductivity as a function of increasing Zeeman energy with fixed $B_{\perp}$.}
%\label{suppfigtwo}
%\end{center}
%\end{figure}

\section*{Phase diagram as a function of the displacement field}

At high electric field the $\nu =0 $ QH state exhibits insulating character $\sigma_{xx} \to 0$. The QH state at high electric field must be layer polarized. For balanced layer or zero electric field the odd parity LL MLG-like bands are anti-symmetric combinations of the top and the bottom layers for both the A and the B sub-lattices, while the even parity BLG-like bands consist of electrons on the B sub-lattice in the middle layer and the symmetric combination of the A sublattices in the top and the bottom layers. At high electric fields, the electrons prefer to occupy the top most layers. In the zeroth LL, the electric field acts like a tunneling term between even and odd parity states.
At low fields, this transition to a layer polarized state from both
% {\bf Allan; What is TCI phase?} {Yafis: I have corrected this to make it consistent with our new nomenclature} 
QPH phase and the QPHF phase is already pre-empted by the edge state localization effects. 
However, in the case of 
%{\bf Allan:  Are we still using this terminology?} {Yafis: Corrected} 
QVF state which is already an insulator, the transition to the another bulk insulating state appears as a peak in the diagonal conductivity and the bulk state is changed. 

\begin{figure}[h]
\begin{center}
\includegraphics[height=2.0in]{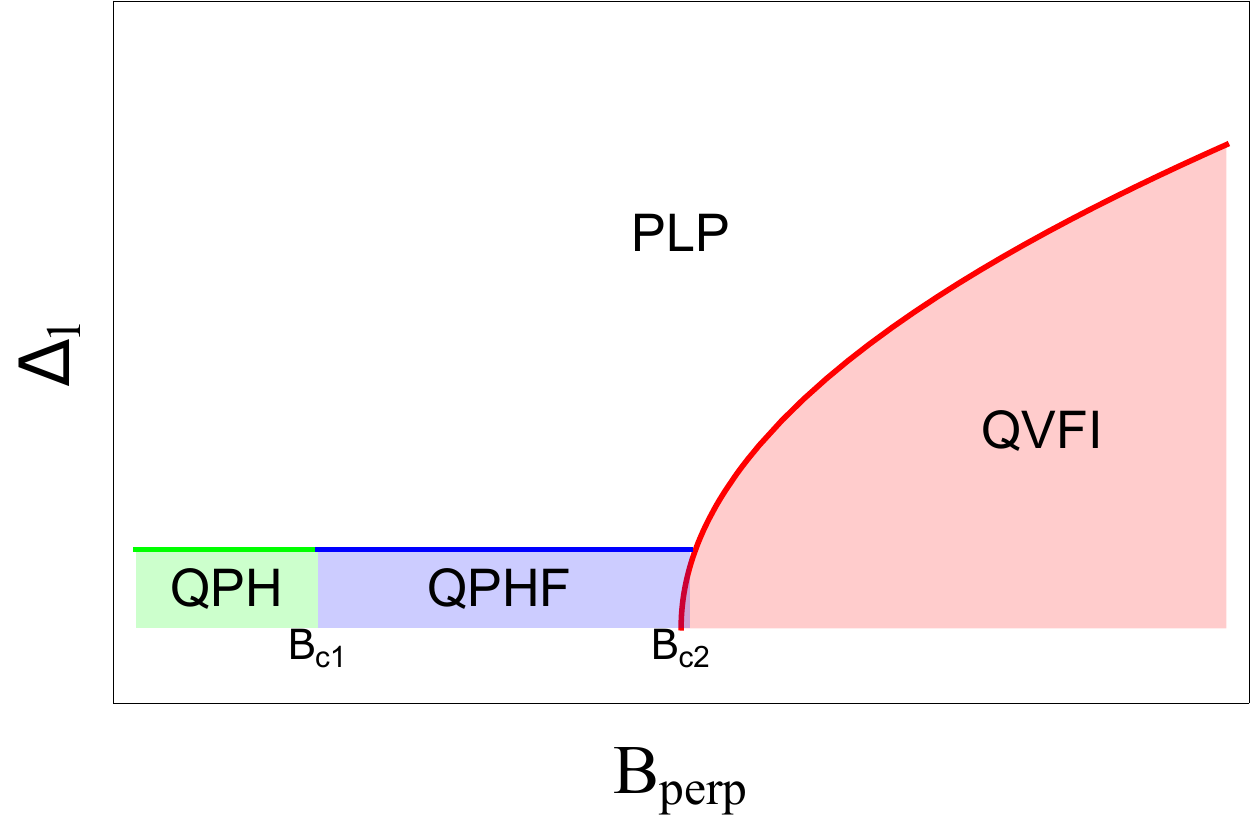}
\caption{Schematic phase diagram of the $\nu =0$ QH state in ABA-trilayer graphene. Phase I is the quantum parity Hall (QPH) state, Phase II is the quantum parity Hall ferromagnetic (QPHF) state, Phase III is the quantum valley ferromagnetic insulator (QVFI) and Phase IV is the partially layer polarized state (PLP) state.
The critical fields for the QPH $\to$ QPHF $\to$ QVFI are denoted by $B_{c1}$ and $B_{c2}$ respectively.}
\label{suppfigthree}
\end{center}
\end{figure}

For $B_{\perp} > B_{c2}$ the $\nu = 0 $ QH state goes through an insulator-insulator transition as a function of the electric field. This bulk transition is due to partial layer polarization and we refer to this state as the partially layer polarized (PLP) state. In the QVF insulating state the reorganization of the even parity electron and odd parity hole LLs occurs due to the interactions. Hence, the incompressible energy gaps due to the interactions scale as square root of the magnetic field. For the layer polarized state energy contribution from the electric field must overcome this interaction energy for a transition to a layer polarized bulk state. Based on the above arguments it is easy anticipate that this phase boundary in the $E$ vs $B_{\perp}$ must scale as $\sqrt{B_{\perp}}$ consistent with the experimental observation. This dependence of the phase boundary can be precisely calculated from out expression of the ground state energy. In the PLP state the electron occupation below the Fermi energy is given by the quantum numbers $(0,A_1,\uparrow ; 0,A_1,\downarrow ; 0,B_2,\uparrow;  0,B_2, \downarrow ; 1,B_2,\uparrow; 1, B_2,\downarrow)$. The ground state energy is given by 
\begin{equation}
\mathcal{E}_{PLP} = -\frac{11}{4} \sqrt{\frac{\pi}{2}} \frac{e^2}{(\epsilon l_{B})} - \frac{\Delta^{B}_{V}}{2} - 2 \Delta_{LL} - 2 \Delta_{mb} - \frac{\Delta^{M}_{V}}{2} - \Delta_{1} + \Sigma^{e}_{0} - 2 \Sigma^e_{1} -  \Sigma^{o}
\end{equation}
Again comparison of the ground state energies $\mathcal{E}_{QVFI} \geq \mathcal{E}_{PLP}$ gives a condition on the self-energies
\begin{equation}
3\Sigma^{e}_{0} + \Sigma^{o} \leq - \sqrt{\frac{\pi}{2}} \frac{e^2}{(\epsilon l_{B})} + \Delta_{mb} + \Delta_{1} -\frac{\Delta^{M}_{V} + 3 \Delta^{B}_{V}}{2},
\end{equation}
which has a strong dependence of the on the exchange interaction. Since we already know that the self-energy has a weak dependence on $B_{\perp}$ we can estimate the phase boundary from the above condition
\begin{equation}
\Delta_{1} \simeq \sqrt{\frac{\pi}{2}} \frac{e^2}{(\epsilon l_{B})} + \Delta^{B}_{V} + 2 \Sigma^e_{0}.
\end{equation}
As anticipated the exchange interaction scaling $\sim \sqrt{B_{\perp}}$ determines the phase boundary of the two competing states. Combining all this one arrives at the schematic phase diagram shown in Fig.~\ref{suppfigthree}

\end{widetext}   


\begin{thebibliography}{}

\bibitem{ref1}	
M. I. Dyakonov, V. I. Perel, Possibility of orienting electron spins with current, Sov. Phys. JETP Lett. {\bf 13}, 467 (1971).

\bibitem{ref2}
M. I. Dyakonov, V. I. Perel, Current-induced spin orientation of electrons in semiconductors, Phys. Lett. A {\bf 35}, 459 (1971).



\bibitem{ref3}
K. F. Mak, K. L. McGill, J. Park, P. L. McEuen, The valley Hall effect in MoS2 transistors, Science {\bf 344}, 1489 (2014).

\bibitem{ref4}
R. V. Gorbachev et al., Detecting topological currents in graphene superlattices, Science, {\bf 346}, 448 (2014).

\bibitem{ref5}
M. Konig et al., The quantum spin Hall effect: Theory and experiment, J. Phys. Soc. Jpn. {\bf 77}, 031007 (2008).
	
\bibitem{ref6}
Z. H. Qiao, W. K. Tse, H. Jiang, Y. G. Yao, Q. Niu, Two-Dimensional Topological Insulator State and Topological Phase Transition in Bilayer Graphene, Phys. Rev. Lett. {\bf 107}, 256801 (2011).


\bibitem{ref7}
J. Ding, Z. H. Qiao, W. X. Feng, Y. G. Yao, Q. Niu, Engineering quantum anomalous/valley Hall states in graphene via metal-atom adsorption: An ab-initio study. Phys. Rev. B {\bf 84}, 195444, (2011).


\bibitem{ref8}
F. Zhang, J. Jung, G. A. Fiete, Q. A. Niu, A. H. MacDonald, Spontaneous quantum Hall states in chirally stacked few-layer graphene systems, Phys. Rev. Lett. {\bf 106}, 156801 (2011).


\bibitem{ref9}
V. M. Apalkov, T. Chakraborty, Electrically tunable charge and spin transitions in Landau levels of interacting Dirac fermions in trilayer graphene, Phys. Rev. B {\bf 86}, 035401 (2012).

\bibitem{ref10}
W. Bao et al., Stacking-dependent band gap and quantum transport in trilayer graphene, Nat Phys {\bf 7}, 948 (2011).

\bibitem{ref11}
C. Cong et al., Raman Characterization of ABA- and ABC-Stacked Trilayer Graphene, ACS Nano {\bf 5}, 8760 (2011).

\bibitem{ref12}
E. A. Henriksen, D. Nandi, J. P. Eisenstein, Quantum Hall Effect and Semimetallic Behavior of Dual-Gated ABA-Stacked Trilayer Graphene, Physical Review X {\bf 2}, 011004 (2012).

\bibitem{ref13}
S. H. Jhang et al., Stacking-order dependent transport properties of trilayer graphene, Physical Review B {\bf 84}, 161408 (2011).

\bibitem{ref14}
M. Koshino, E. McCann, Gate-induced interlayer asymmetry in ABA-stacked trilayer graphene, Physical Review B {\bf 79}, 125443 (2009).

\bibitem{ref15}
Y. Lee et al., Competition between spontaneous symmetry breaking and single-particle gaps in trilayer graphene, Nat Comm. {\bf 5}, 5656, (2014).

\bibitem{ref16}
Y. Liu, S. Goolaup, C. Murapaka, W. S. Lew, S. K. Wong, Effect of Magnetic Field on the Electronic Transport in Trilayer Graphene, ACS Nano {\bf 4}, 7087 (2010).

\bibitem{ref17}
C. H. Lui, Z. Li, K. F. Mak, E. Cappelluti, T. F. Heinz, Observation of an electrically tunable band gap in trilayer graphene, Nat Phys {\bf 7}, 944 (2011).

\bibitem{ref18}
R. Ma, Quantum Hall effect in ABA- and ABC-stacked trilayer graphene, The European Physical Journal B {\bf 86}, 1 (2013).

\bibitem{ref19}
M. Serbyn, D. A. Abanin, New Dirac points and multiple Landau level crossings in biased trilayer graphene, Phys. Rev. B {\bf 87}, 115422 (2013).

\bibitem{ref20}
H. J. van Elferen et al., Fine structure of the lowest Landau level in suspended trilayer graphene, Physical Review B {\bf 88}, 121302 (2013).

\bibitem{ref21}
S. Yuan, R. Roldán, M. I. Katsnelson, Landau level spectrum of \textit{ABA} - and \textit{ABC} -stacked trilayer graphene, Physical Review B {\bf 84}, 125455 (2011).

\bibitem{ref22}
F. Zhang, D. Tilahun, A. H. MacDonald, Hund's rules for the $N=0$ Landau levels of trilayer graphene, Physical Review B {\bf 85}, 165139 (2012).

\bibitem{ref23}
K. Zou, F. Zhang, C. Clapp, A. H. MacDonald, J. Zhu, Transport Studies of Dual-Gated ABC and ABA Trilayer Graphene: Band Gap Opening and Band Structure Tuning in Very Large Perpendicular Electric Fields, Nano Letters {\bf 13}, 369 (2013).

\bibitem{ref24}
Supporting Information

\bibitem{ref25}
W. Zhu, V. Perebeinos, M. Freitag, P. Avouris, Carrier scattering, mobilities, and electrostatic potential in monolayer, bilayer, and trilayer graphene, Physical Review B {\bf 80}, 235402 (2009).

\bibitem{ref26}
M. F. Craciun et al., Trilayer graphene is a semimetal with a gate-tunable band overlap, Nat. Nano. {\bf 4}, 383 (2009).

\bibitem{ref27}
T. Taychatanapat, K. Watanabe, T. Taniguchi, P. Jarillo-Herrero, Quantum Hall effect and Landau-level crossing of Dirac fermions in trilayer graphene, Nat Phys {\bf 7}, 621 (2011).

\bibitem{ref28}
Y. Lee et al., Broken Symmetry Quantum Hall States in Dual-Gated ABA Trilayer Graphene, Nano Letters {\bf 13}, 1627 (2013).

\bibitem{ref29}
W. Bao et al., Magnetoconductance Oscillations and Evidence for Fractional Quantum Hall States in Suspended Bilayer and Trilayer Graphene, Physical Review Letters {\bf 105}, 246601 (2010).

\bibitem{ref30}
P. Stepanov et al., Tunable Symmetries of Integer and Fractional Quantum Hall Phases in Heterostructures with Multiple Dirac Bands, Phys. Rev. Lett. {\bf 117}, 076807 (2016).

\bibitem{ref31}
C. R. Dean et al., Boron nitride substrates for high-quality graphene electronics, Nat Nano {\bf 5}, 722 (2010).

\bibitem{ref32}
T. Taniguchi, K. Watanabe, Synthesis of high-purity boron nitride single crystals under high pressure by using Ba–BN solvent, Journal of Crystal Growth {\bf 303}, 525 (2007).

\bibitem{ref33}
P. Maher et al., Tunable fractional quantum Hall phases in bilayer graphene, Science {\bf 345}, 61 (2014).

\bibitem{ref34}
B. Datta et al., Strong electronic interaction and multiple quantum Hall ferromagnetic phases in trilayer graphene, Nature Communications {\bf 8}, 14518 (2017).

\bibitem{ref35}
J. González, Electron self-energy effects on chiral symmetry breaking in graphene, Physical Review B {\bf 85}, 085420 (2012).
	
\bibitem{ref36}
J. H. a. J. P. F. L. a. J. P. Carbotte, Optical self-energy in graphene due to correlations, Journal of Physics: Condensed Matter {\bf 24}, 245601 (2012).

\bibitem{ref37}
D. A. Siegel et al., Many-body interactions in quasi-freestanding graphene, Proceedings of the National Academy of Sciences {\bf 108}, 11365 (2011).

\bibitem{ref38}
A. F. Young et al., Tunable symmetry breaking and helical edge transport in a graphene quantum spin Hall state, Nature {\bf 505}, 528 (2014).

\bibitem{ref39}
P. Maher et al., Evidence for a spin phase transition at charge neutrality in bilayer graphene, Nat Phys {\bf 9}, 154 (2013).

\bibitem{ref40}
Yafis Barlas, Counter-propagating Fractional Hall states in mirror-symmetric Dirac semi-metals, Phys. Rev. Lett. {\bf 121}, 066602 (2018). 


\end{thebibliography}

\begin{thebibliography}{99}

\bibitem{grapheneQHreview}
Barlas Y., Yang K. and MacDonald A. H. Quantum Hall effects in graphene-based two-dimensional electron systems, {\it Nanotechnology} {\bf 23}, 052001 (2012).

\bibitem{AbaninABA}
Serbyn M. and Abanin D. A. New Dirac points and multiple Landau level crossings in biased trilayer graphene, {\it Phys. Rev. B} {\bf 87}, 115422 (2013).

\end{thebibliography}
\end{document}